# Playing catch-up in building an open research commons
*As efforts advance around the globe, the US falls behind*

By Philip E. Bourne1*, Vivien Bonazzi2, Amy Brand3, Bonnie Carroll4, Ian Foster5, Ramanathan V. Guha6, Robert Hanisch7, Sallie Ann Keller8, Mary Lee Kennedy9, Christine Kirkpatrick10, Barend Mons11, Sarah M. Nusser12, Michael Stebbins13, George Strawn14, and Alex Szalay15
1School of Data Science, University of Virginia; Charlottesville, VA, USA.2Deloitte; Washington DC, USA. 3MIT Press; Cambridge, MA, USA.4Information International Associates; Oak Ridge, TN, 37830, USA.5Argonne National Laboratory; Lemont, IL, USA.6Google; Mountain View, CA, USA.7Office of Data and Informatics, National Institute of Standards and Technology (NIST); Gaithersburg, MD, USA.8Biocomplexity Institute, University of Virginia; Charlottesville, VA, USA.9Association of Research Libraries; Washington DC, USA.10San Diego Supercomputer Center; San Diego, CA, USA.11Human Genetics Department, and Leiden Academic Center for Drug Research Leiden University; Leiden, The Netherlands.12Department of Statistics, Iowa State University; Ames, IA, USA.13Science Advisors LLC, Alexandria, VA USA.14Vienna, VA, USA.15Department of Physics and Astronomy, John Hopkins University; Baltimore, MD, USA.
Email: peb6a@virginia.edu

Many aspects of the research enterprise are rapidly changing to be more open, accessible, and supportive of rapid-response investigations (e.g., understanding COVID) and large cross-national research that addresses complex challenges (e.g., supply chain issues). Around the globe, there have been aggressive responses to the need for a unified open research commons (ORC) - an interoperable collection of data and compute resources within both the public and private sectors which are easy to use and accessible to all. Many nations are positioning themselves to be scientifically competitive in years to come. But the United States is falling behind in the accessibility and connectedness of its research computing and data infrastructure (1), compromising competitiveness and leadership and limiting global science that could benefit from US contributions. The challenge is more cultural and institutional than technical, demanding immediate and sustained leadership and support, starting with policy makers and research funders.

The value of cooperation around technology and data broadly is beyond question. For example, shared governance, shared infrastructure, and agreements on standards that permit a shared system to operate underpin the ability of the North American electrical grid to direct electricity where it is needed, and of the CIRRUS banking network to deliver money from one's bank to almost anywhere in the world. The sum of the parts is made greater than the whole through cooperation across the enterprise.

Similar coordination in the research enterprise can pay enormous dividends. We now have vast amounts of publicly available research data, but to fully leverage the potential power of these data beyond individual and often heroic efforts, these data need to be identified, made interoperable, and aligned so they can be broadly utilized by the scientific community. For example, often data on disparate topics – e.g., a county's homelessness rates, average income, neighborhood food and health resources, air pollution, flood risk, predicted water resources, and predicted average temperature – are spread across a range of locations on the web, infrastructures, and management regimes.    If these data were integrated (bought together based on common data elements in each data set), we could use it for powerful analyses, like identifying locations with high homeless populations that are also likely to be hit hardest by floods, droughts, or heat waves; or places with poor cardiac health that also have high or increasing PM2.5 pollution, which could lead to more heart attacks. Support by policy makers and funders who are driving the development of research infrastructure can facilitate such work, similar to the urgent cooperation we see among scientists during a time of dire need, like the COVID pandemic, the threat of war, and the disruption to the global economy.

In principle, this should be possible in the US, which has a vibrant research ecosystem with no lack of computation and data resources. But establishing an ORC is less a technical challenge than it is a cultural and institutional one that requires policy leadership and a sustained commitment, both of which have been lacking. Talk to any of the burgeoning number of data scientists and they will likely tell you about, in accessing data that are already publicly available , all the time they must spend learning a variety of esoteric compute systems, figuring out where to get data and what value the data have before they even begin the real work of analysis and discovery. Common application programming interfaces (API's), new metadata and data standards, workflows (2) , dashboards and evaluation of progress as these evolve would enable them to better solve the problems facing society. Beyond the technical aspects, there are also     social/cultural aspects of appropriate recognition that data are first class research objects that should be cited and acknowledged in the same way that publications are today.

In recognition of these challenges and opportunities, in 2013, the White House Office of Science and Technology Policy (OSTP) issued a directive to Federal agencies to improve access to the results of scientific research outcomes. The intentions of the Administration were made clear: "digitally formatted scientific data resulting from unclassified research supported wholly or in part by Federal funding should be stored and publicly accessible to search, retrieve, and analyze." This US initiative was an impetus to the global push to open government, open science, and open data.

Countries around the world have since forged ahead. A wide range of recent efforts reflect substantial recognition in other countries and regions of the important role played by ORCs. For example: the European Open Science Cloud (EOSC), the CS3MESH4EOSC Science Mesh, the China Science and Technology (CST) Cloud, the African Open Science Platform (AOSP), the South African National Integrated Cyber Infrastructure System (NICIS), the Malaysia Open Science Platform (MOSP), the Global Open Science Cloud (GOSC) funded mainly by China and organized through the International Science Council's Committee on Data (CODATA), the Australian Research Data Commons (ARDC) Nectar Research Cloud, the Digital Research Alliance of Canada (formerly known as the New Digital Research Infrastructure Organization, NDRIO), and the Arab States Research and Education Network (ASREN).

Although all are quite recent initiatives, already the CST Cloud serves over one million researchers. ARDC is supporting studies of bushfires, which have plagued Australia in recent years, and is working with all 43 universities across Australia to build a nationally-agreed upon network for research data. The EOSC-related projects and the European Member States participate with considerable additional investments governed through the EOSC association spanning the full scope of scientific disciplines, and provides support to high-performance computing centers, massive databases, and the software tools required to utilize them. As a result, EOSC is pivotal in supporting research challenges such as climate change, space weather, seismology, bioinformatics, disaster mitigation, toxicology, and radio astronomy. Notwithstanding such early signs of progress, systematic evaluation of these developing projects is needed to assess their full value.

Critically, the governance models for many of these programs seek to enable equitable access to research capacity, consonant with a major policy goal of the current US administration. For example, substantial effort, as stated in the EOSC Declaration, went into defining rules of participation, processes for governance, and allocation of resources.

FRAGMENTED, FALLING BEHIND
As these collaborative initiatives advance around the globe, anecdotally progress in establishing an ORC has seemed to fall behind in the US because of a vacuum of leadership, focus, and coordination. Without a coherent national strategy, US scientists have found it harder to participate in the broader global pursuit of ORCs, eroding US competitiveness down to how it impacts individual scientists.

The major US research funding agencies largely tend to pursue independent initiatives and rarely work together on shared infrastructure, even as the problems society faces span agencies. Aside from a few shared resources such the National Science Foundation's Extreme Science and Engineering Discovery Environment (XSEDE), with its focus on the use of high-performance computing, and Open Science Grid (OSG) Consortium, with shared compute and data resources (mostly NSF-sponsored), we assert that U.S. research computational and data infrastructure remains fragmented, inefficient, and uncompetitive with ORC's that are emerging globally.

Funding would seem to be diminishing, as in the case of XSEDE's follow on program, Advanced Cyberinfrastructure Coordination Ecosystem: Services & Support (ACCESS), which is resourced to provide roughly half of the funds of XSEDE, despite increased usage and reliance of US scientists on advanced computing. The OSG Consortium, in contrast, is more robust with a distributed funding model, yet a single administrative council and software support services. At the same time, there is little equity to the resources that exist. Access tends to favor the major research-intensive universities where substantial expertise and local compute access already exist. More can be achieved by pooling publicly available data and compute resources and enabling a larger and more diverse group of researchers, who are from underrepresented communities and found in universities and elsewhere, to easily access those resources to address society's most pressing problems.

The tech industry is working to provide some integration of and access to data, but it is often not done with proper research context and is driven by profit potential, not scientific need. The research community should embrace US industry potential where mutual benefit can be found, thus partnering and building trust rather than be dependent. Yet public-private partnerships (PPP) are few and limited by government rules, leading to separate organizations such as the Foundation for the NIH (FNIH) which facilitates joint initiatives between NIH and the private sector in ways not allowed by a federal agency alone. Hence, the interdisciplinary knowledge, innovation, and underlying shared data and compute resources needed to solve global challenges are usually lacking.

The lack of a unified ORC leaves the US following along in global efforts (e.g., GOSC) with no formal executive representation in international ORC-focused initiatives such as at the Open Science Clouds and Commons Executives' Roundtable (OSCER), where executives of many of the international initiatives collaborate in creating an interconnected global research cloud infrastructure built from the individual research commons developed by other nations.

We are encouraged by recent efforts to improve access to computation and data infrastructure, for example, in the National Artificial Intelligence Initiative. However, this represents a partial and siloed effort to serve specific communities largely focused on machine learning and artificial intelligence (AI) applications. While important, there is a larger research ecosystem that needs access to data and computational resources.

Data, in particular, need to be made AI-ready, preserved for substantial periods of time, in support of reproducibility, for ensuring the integrity of the scholarly research record, and either because they cost so much to produce as to be irreplaceable and/or the aggregation of such data may lead to new insights at some indeterminate future time. But the long-term beneficiaries may be very different from the producers or initial custodians, which adds to the challenge and needs to be part of a national ORC concept. This requires professional data stewardship, a skill that is strongly developed in the various regional initiatives around the globe and exemplified by the International Society for Biocuration.

ALIGNING INCENTIVES

It is imperative that together all stakeholders create a more seamlessly connected and accessible data and compute infrastructure. What is needed is commitment from the Congress and the Administration so parties can come together to chart the future and address the fragmented nature of the US research computing and data enterprise by establishing an Open Research Commons. Building on prior conceptualizations (3,4), the ORC should span federal, state, and local government agencies and computing facilities, including national laboratories, public and private clouds and institutions. The ORC should be designed to replace the largely siloed, individually controlled data and compute resources in the US today, a situation that limits discoverability, access, innovation and collaboration.

This will require shared governance, trust, common standards, and shared infrastructure and most importantly a champion, such as the Office of Science and Technology Policy (OSTP). A key to success is the incentive to create a unified system. Scientists are not yet presented with the adequate incentives. Mandates from funders, such as data sharing policies, help, but there is not enough definition of requirements, reward for complying, and indeed a unification of what is expected of researchers regardless of the source of their research funding. What is lacking is the coordination and incentives to establish shared data and compute resources, governance, policies, and procedures across the US research enterprise to maximize accessibility and equity and train a computation- and data-savvy workforce.

Beyond being competitive is the need to contribute to the global ecosystem such that beyond the will to collaborate is the ORC infrastructure to make it productive to do so. Efforts in the US must thus absolutely learn from and engage with others abroad, to adopt and adapt rather than start from scratch and reinvent what is already known. Efforts must reach out across sectors. Though commercial entities must be part of the ecosystem, they alone cannot be expected to adequately develop an ORC that serves the public good and needs of scientists, particularly as a long-term commitment. An ORC would thus provide a focal point for creating and nurturing PPPs by adopting shared research data policies, standards and practices as well as the technical infrastructure to reduce the risk of absent, inoperable and lost knowledge at the time of need.

Much of the hardware, software and knowledge needed to make the ORC a reality largely exists.    Thus, the cost may not be overly daunting compared to the gains in productivity, insight, economic competitiveness, and security that may emerge. While the exact return on investment is hard to estimate (more explicit collaborative research is needed here), an increase in research productivity at even the 10% level would likely far outweigh the cost of establishing the ORC.

Recent developments in the US point to expected benefits of a unified and coordinated approach. Notable are the NSF's Research Coordination Network (RCN) and the NSF program of investment in a Research Data Ecosystem: A National Resource for Reproducible, Robust, and Transparent Social Science Research in the 21st Century. But much more is needed.

The Congress and Administration must take action now . That action should be defined by the     OSTP    and include mandates to all federally funded science and technology agencies. Mandates for cooperation across agencies, like that currently being attempted by the Global Biodata Coalition (GBC), albeit more focused on sustainability, that lead to shared compute infrastructure and, if not shared, governed by a set of rules and standards that facilitate data exchange and reuse and provide a consistent interface to both humans and machines. Mandates that have the agencies working in closer cooperation with the private sector to realize the full potential of the US technology workforce without compromising competition. Only then will there be     opportunity to maximize productivity and innovation to solve global problems    . Other countries and regions are taking these same steps. It is time for the US to step up to the plate.

ACKNOWLEDGEMENTS
This commentary draws upon discussions held at an event, Towards a Unified US Open Research Commons, held on 2 August 2021  in Washington DC,  with support from the N ational Academies of Science, Engineering and Medicine. The event     was inspired in part by (1) . These opinions, recommendations, findings, and conclusions do not necessarily reflect the views or policies of NIST or the United States Government.